\documentclass[reprint,aip,jap]{revtex4-1}

\usepackage{graphicx}
\usepackage{dcolumn}
\usepackage{bm}
\usepackage{color}

\begin{document}

\preprint{Appl.~Phys.~Lett.~(2012)}

\title{Spin transport and spin dephasing in zinc oxide}

\author{Matthias Althammer}
    \affiliation{Walther-Mei{\ss}ner-Institut, Bayerische Akademie der Wissenschaften, 85748 Garching, Germany}
\author{Eva-Maria Karrer-M\"{u}ller}
    \affiliation{Walther-Mei{\ss}ner-Institut, Bayerische Akademie der Wissenschaften, 85748 Garching, Germany}
\author{Sebastian T. B. Goennenwein}
    \affiliation{Walther-Mei{\ss}ner-Institut, Bayerische Akademie der Wissenschaften, 85748 Garching, Germany}
\author{Matthias Opel}
    \email{Matthias.Opel@wmi.badw.de}
    \affiliation{Walther-Mei{\ss}ner-Institut, Bayerische Akademie der Wissenschaften, 85748 Garching, Germany}
\author{Rudolf Gross}
    \affiliation{Walther-Mei{\ss}ner-Institut, Bayerische Akademie der Wissenschaften, 85748 Garching, Germany}
    \affiliation{Physik-Department, Technische Universit\"{a}t M\"{u}nchen, 85748 Garching, Germany}

\date{\today}

\begin{abstract} 
The wide bandgap semiconductor ZnO is interesting for spintronic applications because of its small spin-orbit coupling implying a large spin coherence length. Utilizing vertical spin valve devices with ferromagnetic electrodes (TiN/Co/ZnO/Ni/Au), we study the spin-polarized transport across ZnO in all-electrical experiments. The measured magnetoresistance agrees well with the prediction of a two spin channel model with spin-dependent interface resistance. Fitting the data yields spin diffusion lengths of 10.8\,nm (2\,K), 10.7\,nm (10\,K), and 6.2\,nm (200\,K) in ZnO, corresponding to spin lifetimes of 2.6\,ns (2\,K), 2.0\,ns (10\,K), and 31\,ps (200\,K).
\end{abstract}


\maketitle


The successful injection, transport, manipulation, and detection of spin-polarized currents in semiconductors is a prerequisite for semiconductor spintronics.\cite{Prinz1998} In this context, the spin dephasing time $T_2^\star$ of mobile charge carriers -- and the associated length scale for coherent spin transport -- are fundamental parameters. The wide bandgap II-VI semiconductor zinc oxide has a small spin-orbit coupling \cite{Fu2008} implying a large spin coherence length. This makes ZnO interesting for (opto)electronics or spin-based quantum information processing. While other semiconductors like GaAs and related III-V compounds have been studied extensively,\cite{Awschalom2002} only few reports on spin-coherent properties in ZnO exist.\cite{Ghosh2005,Liu2007,Bratschitsch2008} Using time-resolved Faraday rotation (TRFR), electron spin coherence up to room temperature in epitaxial ZnO thin films was first observed by Ghosh \textit{et al.} with a spin dephasing time of $T_2^\star \simeq 2$~ns at 10\,K.\cite{Ghosh2005} Our ZnO thin films display $T_2^\star \simeq 14$\,ns (also at 10\,K) in similar optical experiments.\cite{Weier2010,Opel2012} Reports on electrical spin injection, however, are rare\cite{Chen2002,Ji2009,Shimazawa2010} and mainly focus on technical aspects rather than fundamental spin-dependent properties of ZnO. The highest reported values for the magnetoresistance (MR) in ZnO-based spin valve structures are 1.38\% and 1.12\% (90\,K) for thicknesses of the ZnO layer of 3\,nm and 10\,nm, respectively.\cite{Ji2009} Here, we investigate the transport and the dephasing of spin-polarized charge carriers in ZnO utilizing all-electrical, vertical spin valve devices with ferromagnetic (FM) electrodes. We do not focus on the mechanism of spin injection and the general problem of overcoming the conductance mismatch\cite{Schmidt2000} between ZnO and the ferromagnetic layers in the spin valve structure. The experimental observation of a large MR\cite{Baibich1988,Binasch1989,Gijs1997} of up to 8.4\% in our FM/ZnO/FM junctions demonstrates the successful transport of a spin-polarized ensemble of electrons across several nanometers in ZnO.

\begin{figure}
    \includegraphics[width=\columnwidth]{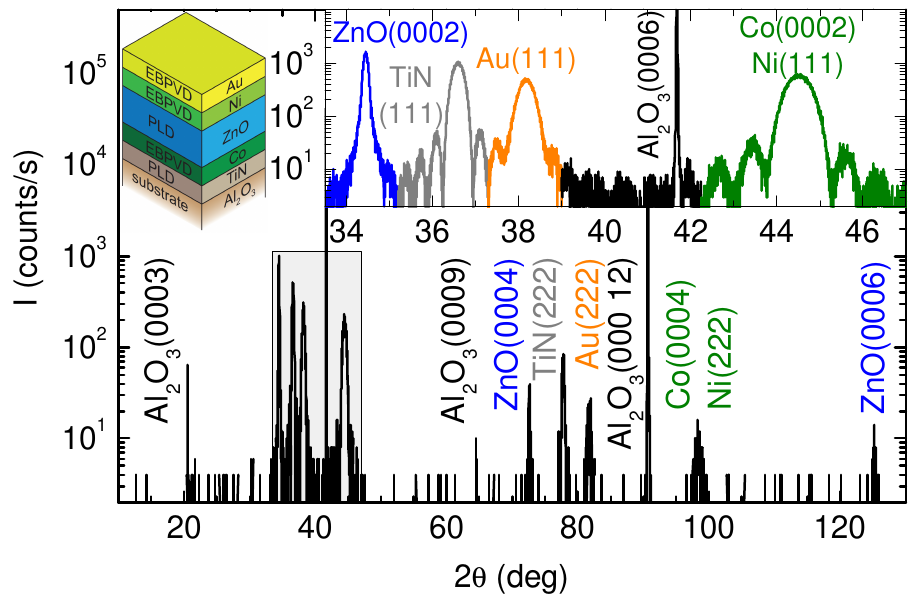} 
    \caption{\label{fig:XRD} High-resolution X-ray diffraction diagram from a TiN/Co/ZnO(80\,nm)/Ni/Au multilayer stack, deposited by pulsed laser deposition (PLD) or electron beam physical vapor deposition (EBPVD). The $\omega$-$2\theta$-scan does not show any secondary phases. The shaded area around the Al$_2$O$_3$(0006) reflection is displayed in the right inset on an enlarged scale. Satellites due to Laue oscillations are clearly resolved for TiN(111) and Co(0002).}
\end{figure}

The thin film multilayer heterostructures (left inset in Fig.\,\ref{fig:XRD}) were fabricated on (0001)-oriented, single crystalline Al$_2$O$_3$ substrates via laser-MBE in combination with electron-beam physical vapor deposition (EBPVD) in an ultra-high vacuum cluster system.\cite{Gross2000} Laser-MBE was carried out by pulsed laser deposition (PLD) from stoichiometric polycrystalline targets, using a KrF excimer laser with a wavelength of 248\,nm at a repetition rate of 10\,Hz.\cite{Opel2012} EBPVD was performed in high vacuum ($<\!5\!\times\!10^{-7}$\,mbar) at room temperature. The multilayer stack consists of (i) a 12\,nm thin TiN film as a non-ferromagnetic, metallic bottom electrode, deposited via PLD at $600^\circ$C in Ar atmosphere of $7\times10^{-4}$\,mbar with a laser fluence of 2\,J/cm$^2$; (ii) a 11\,nm thin Co film (EBPVD) as the first ferromagnetic electrode; (iii) a semiconducting ZnO film (PLD, $400^\circ$C, Ar, $1\times10^{-3}$\,mbar, 1\,J/cm$^2$) with thickness $t_\mathrm{ZnO}$ systematically varied in the range from 5\,nm to 100\,nm in a series of samples; (iv) a 11\,nm thin Ni film (EBPVD) as the second ferromagnetic electrode; and finally (v) a 24\,nm thin Au film (EBPVD) as a capping layer and top electrode.

The structural quality of the multilayer stack was investigated by high-resolution X-ray diffractometry in a four circle diffractometer using Cu K$\alpha_1$ radiation. The out-of-plane $\omega$-$2\theta$-scan does not reveal any secondary phases (Fig.\,\ref{fig:XRD}). All detected reflections can be assigned to the respective materials of the spin valve multilayer structure. The cubic materials TiN, Ni, and Au grow (111)-oriented, the hexagonal Co and ZnO layers (0001)-oriented. The signals from Co and Ni cannot be separated as the $c$ lattice parameter of hcp Co ($c_{\mathrm Co} = 0.407$\,nm)\cite{Martienssen2005} is twice as large as the lattice spacing $d_{111} = \frac{1}{3} \sqrt{3} a = 0.203$\,nm of the $[111]$-planes in cubic Ni with $a_{\mathrm Ni} = 0.352$\,nm.\cite{Martienssen2005} For the TiN, Co, and Au layers, the $\omega$-$2\theta$-scan shows satellites due to Laue oscillations (right inset of Fig.\,\ref{fig:XRD}) demonstrating coherent growth with small interface roughness and indicating a high structural quality. From the out-of-plane reflections, we calculate the corresponding lattice parameters for each layer ($a_\mathrm{TiN}=0.425$\,nm, $c_\mathrm{Co}=0.407$\,nm, $c_\mathrm{ZnO}=0.521$\,nm, $a_\mathrm{Ni}=0.352$\,nm, $a_\mathrm{Au}=0.408$\,nm) and find them very close ($<0.1\%$) to the respective bulk values.\cite{Martienssen2005} The high structural quality is also confirmed by the narrow full widths at half maximum (FWHM) of the $\omega$ rocking curves (not shown here) for the respective out-of-plane reflections. We obtain $\mathrm{FWHM}=0.03^\circ$ for TiN(111), $0.04^\circ$ for Co(0002), $0.62^\circ$ for ZnO(0002), and $1.47^\circ$ for Au(111), demonstrating a low mosaic spread. $\phi$-scans around asymmetric reflections (not shown here) exhibit a sixfold in-plane symmetry for all layers with clear epitaxial relationships: Al$_2$O$_3(0001) [11\overline{2}0] \parallel$ TiN$(111) [2\overline{1}\overline{1}] \parallel$ Co$(0001) [10\overline{1}0] \parallel$ ZnO$(0001) [10\overline{1}0] \parallel$ Ni$(111) [2\overline{1}\overline{1}] \parallel$ Au$(111) [2\overline{1}\overline{1}]$. From this detailed structural analysis, it becomes evident that each layer is aligned with respect to the oxygen sublattice of the (0001)-oriented Al$_2$O$_3$ substrate. This effect has already been reported for TiN(111),\cite{Talyansky1999} Co(0001),\cite{Ago2010} ZnO(0001),\cite{Chen1998} and Au(111).\cite{Kastle2002} Our results prove that the respective in-plane orientations of the single layers are preserved with excellent quality when grown on top of each other.

\begin{figure}
    \includegraphics[width=\columnwidth]{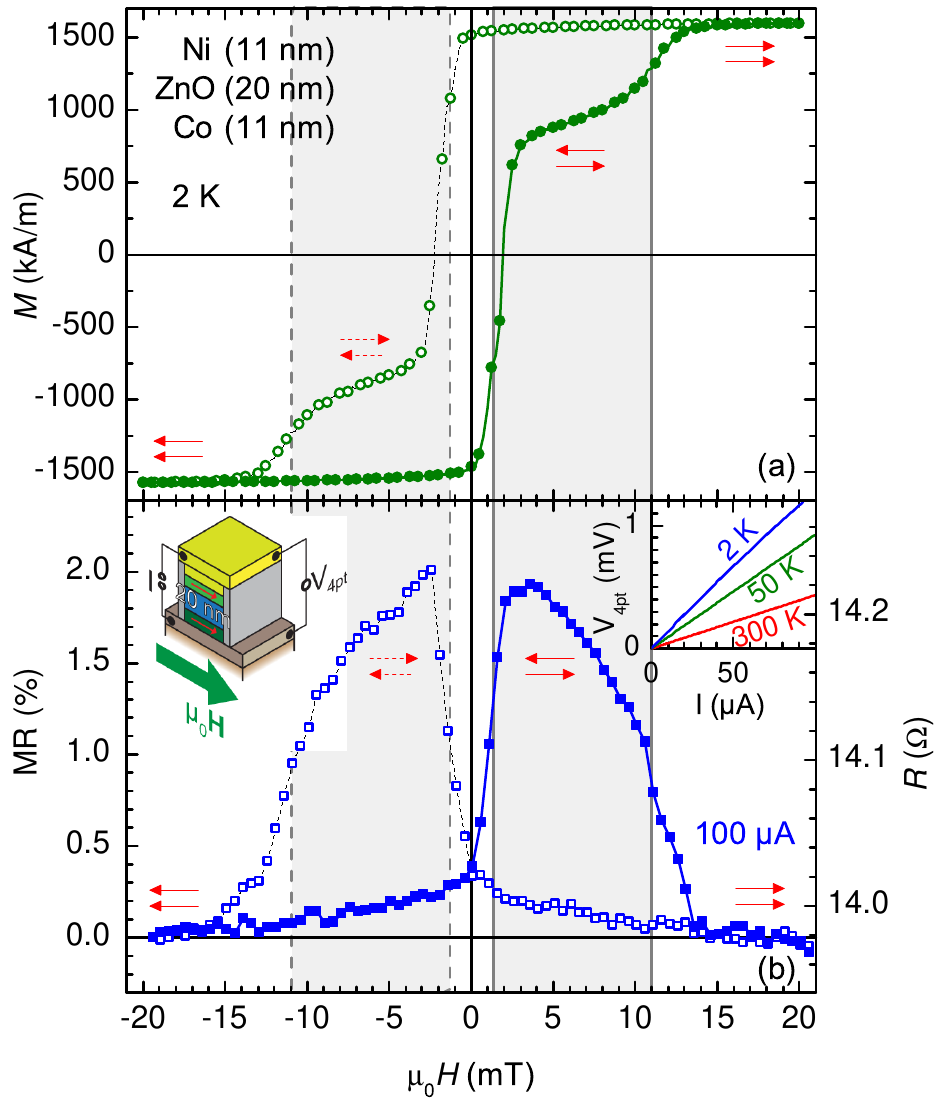} 
    \caption{\label{fig:SQUID-MR} (a) Magnetization $M$ of the multilayer stack normalized to the total volume of both ferromagnetic layers and (b) magnetoresistance MR for the patterned spin valve device (left inset) with an area of $400\,\mu\mathrm{m}^2$. The data were taken for a ZnO thickness of 20\,nm at 2\,K as a function of the in-plane magnetic field $H$ for up (closed symbols) and down sweep (open symbols). The MR effect is correlated to the coercive fields of Co and Ni (vertical lines); its maximum of $1.93\%$ appears in the regime where the magnetizations of Co and Ni (red arrows) are aligned antiparallel (grey shaded areas). In (b), the right scale shows the absolute four-point resistance and the right inset the $I$-$V$-characteristics of the device.}
\end{figure}

The magnetic properties of the multilayer stack were studied via superconducting quantum interference device (SQUID) magnetometry at temperatures between 2\,K and 300\,K in magnetic fields of $|\mu_0 H| \leq 7$\,T applied in-plane. At all investigated temperatures, the magnetization $M$ shows a ferromagnetic hysteresis for low fields. The shape of $M(H)$ represents a superposition of two distinct hysteresis curves with different coercive fields $H_\mathrm{c}$ (Fig.\,\ref{fig:SQUID-MR}(a)). We assign the larger $H_\mathrm{c}$ to the Ni film and the lower $H_\mathrm{c}$ to the Co layer, as confirmed by investigating a modified structure without Ni (not shown here). From these coercivities, the magnetizations of Ni and Co (red horizontal arrows in Fig.\,\ref{fig:SQUID-MR}) are expected to be aligned parallel for $|H| \gg H_\mathrm{c}^\mathrm{Ni},H_\mathrm{c}^\mathrm{Co}$ and antiparallel (shaded regions in Fig.\,\ref{fig:SQUID-MR}) for $H_\mathrm{c}^\mathrm{Co} < H < H_\mathrm{c}^\mathrm{Ni}$ for the field up-sweep direction or $-H_\mathrm{c}^\mathrm{Co} > H > -H_\mathrm{c}^\mathrm{Ni}$ for the sweep-down direction, respectively. This independent switching of the magnetizations is a key requirement for the successful realization of a spin valve device.

Using photolithography, Ar ion beam milling and lift-off processes, the multilayer stack was patterned into vertical mesa structures with junction areas of $400\,\mu$m$^2$. These spin valve junctions were investigated by magnetotransport at temperatures $T$ between 2\,K and 300\,K in a liquid He magnet cryostat system with the magnetic field $H$ again applied in-plane. The vertical transport was studied in standard four-point geometry by applying a constant dc bias current $I$ and measuring the voltage drop $V_\mathrm{4pt}$ across the spin valve junction (left inset of Fig.\,\ref{fig:SQUID-MR}(b)). The $I$-$V$-characteristics of all investigated samples display a linear regime for small bias currents of $I \lesssim 100\,\mu$A (right inset in Fig.\,\ref{fig:SQUID-MR}(b)).\footnote{The linear behavior is expected in the tunneling regime. For higher currents, heating leads to non-linear $I$-$V$-characteristics.} In the following and for the evaluation of the magnetoresistance (MR) of samples with different thicknesses $t_\mathrm{ZnO}$ of the ZnO layer below, we restrict $I$ to this Ohmic regime for all investigated temperatures.

\begin{figure}
    \includegraphics[width=\columnwidth]{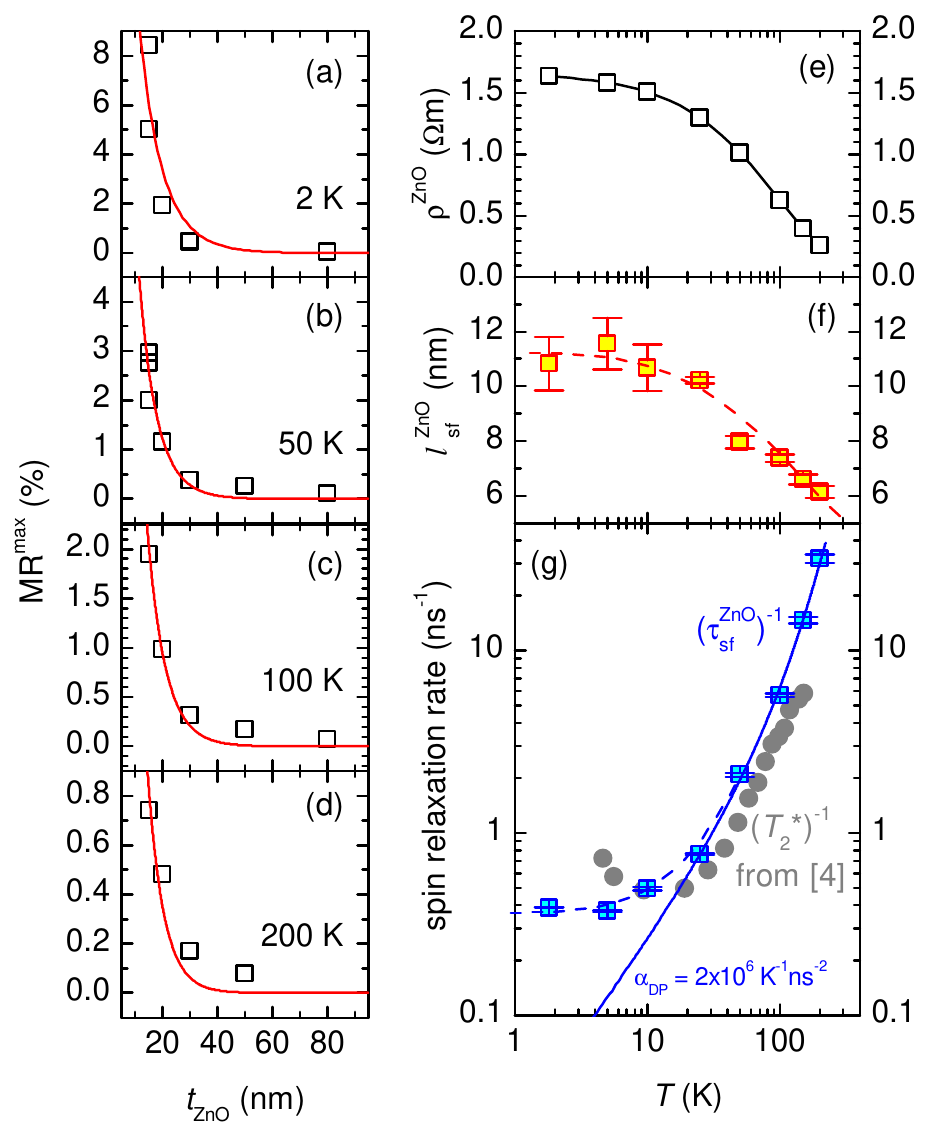} 
    \caption{\label{fig:fitting} (a--d) Maximum MR as a function of the ZnO thickness $t_\mathrm{ZnO}$ at 2\,K (a), 50\,K (b), 100\,K (c), and 200\,K (d). The symbols represent the experimental data, the red lines are fits with the two spin channel model.\cite{Valet1993,Fert2001} (e) Temperature dependence of the resistivity $\rho_\mathrm{ZnO}$, experimentally determined from a TiN/ZnO/Au reference sample. (f) Spin diffusion length $l_\mathrm{sf}^\mathrm{ZnO}$, obtained from the fits in (a--d). The dashed line is a guide to the eye. (g) Inverse spin lifetime $\tau_\mathrm{sf}^\mathrm{ZnO}$ (blue squares), obtained from $l_\mathrm{sf}^\mathrm{ZnO}$ via Eq.\,(\ref{eq:tau}). The solid line corresponds to Eq.\,(\ref{eq:DP}) with $\alpha_\mathrm{DP} = 2 \times 10^6\,\mathrm{K^{-1}ns^{-2}}$, the dashed line is a guide to the eye. The panel also shows $T_2^\star$ data (grey circles), determined optically by Ghosh \textit{et al.}\cite{Ghosh2005}}
\end{figure}

The resistance $R = V_\mathrm{4pt}/I$ is found to scale with $t_\mathrm{ZnO}$, demonstrating that it is not dominated by the ferromagnetic layers. The resistivity $\rho(T)$ determined by multiplying $R$ with the junction area and dividing by the respective thickness of the whole multilayer stack shows the same temperature dependence for all spin valve junctions as for a reference junction of TiN/ZnO/Au without ferromagnetic electrodes. $\rho(T)$ increases by about one order of magnitude from 300\,K to 2\,K (Fig.\,\ref{fig:fitting}(e)). From the analysis of the Hall effect of our ZnO thin films deposited directly on Al$_2$O$_3$ substrates, we obtained a high and almost constant carrier concentration of $n^\mathrm{ZnO} \simeq 5 \times 10^{17}\,\mathrm{cm}^{-3}$ over the whole temperature range (not shown here). Assuming an effective mass of $m^\star = 0.3 m_\mathrm{e}$, from this value we estimate a Fermi temperature of $T_\mathrm{F} = 89$\,K. With $\rho(T)$ from Fig.\,\ref{fig:fitting}(e), we further find a low carrier mobility of $\mu^\mathrm{ZnO}(T) \lesssim 1\,\mathrm{cm^2/Vs}$.

When sweeping the magnetic field, $R(H)$ shows a hysteresis between the field up-sweep and down-sweep directions with two resistive states (Fig.\,\ref{fig:SQUID-MR}(b)), corresponding very well to the $M(H)$ hysteresis discussed above. For magnetic fields with an antiparallel magnetization configuration of the Ni and Co electrodes (shaded regions in Fig.\,\ref{fig:SQUID-MR}), $R$ is significantly higher than for the parallel configurations. This is evidence that our vertical mesa structures act as spin valve devices.\footnote{This statement is further confirmed by angle-resolved measurements, see supplemental material.} For $t_\mathrm{ZnO} = 20$\,nm and at 2\,K, the magnetoresistance
\begin{equation}
    \mathrm{MR} = \frac{R(H)-R(-200\,\mathrm{mT})}{R(-200\,\mathrm{mT})}
\end{equation}
reaches a maximum value MR$^\mathrm{max}$ of almost 2\% (Fig.\,\ref{fig:SQUID-MR}(b)). MR$(H)$ is independent of the potential difference across the spin valve for dc bias currents up to $I = 100\,\mu$A. Evaluating MR$(H)$ from different samples with different thicknesses $t_\mathrm{ZnO} = 15 \ldots 80$\,nm of the semiconducting ZnO layer, we find that MR$^\mathrm{max}$ sensitively depends on $t_\mathrm{ZnO}$. At 2\,K, MR$^\mathrm{max}$ rapidly decreases from 8.4\% for 15\,nm to 0.06\% for 80\,nm (Fig.\,\ref{fig:fitting}(a)). The same behavior is observed at higher temperatures of 50\,K, 100\,K, and 200\,K with the overall values becoming significantly smaller (Fig.\,\ref{fig:fitting}(b-d)).


Following the Valet-Fert approach,\cite{Valet1993,Fert2001} we fit MR$^\mathrm{max}$ to a two spin channel model with spin-dependent conductivities to extract the spin diffusion length $l_\mathrm{sf}^\mathrm{ZnO}(T)$ for ZnO. We use the same nomenclature as in Ref.~\onlinecite{Fert2001}. As material parameters we plug in the resistivities $\rho^\mathrm{Ni} = 59$\,n$\Omega$m and $\rho^\mathrm{Co} = 56$\,n$\Omega$m,\cite{Martienssen2005} the spin asymmetry coefficients $\beta^\mathrm{Ni} = -0.14$ (Ref.~\onlinecite{Moreau2007}) and $\beta^\mathrm{Co}=-0.46$ (Ref.~\onlinecite{Bass1999}), as well as the spin diffusion lengths $l_\mathrm{sf}^\mathrm{Ni} = 21$\,nm (Ref.~\onlinecite{Moreau2007}) and $l_\mathrm{sf}^\mathrm{Co} = 59$\,nm (Ref.~\onlinecite{Piraux1998}) for Ni and Co, respectively. Since these parameters have only small influence on the MR, we assume for simplicity that they are independent of temperature. The spin-dependent resistivities of Ni and Co are then given by $\rho^\mathrm{Ni,Co}_{\uparrow (\downarrow)} = 2\rho^\mathrm{Ni,Co} [1 -(+) \beta^\mathrm{Ni,Co}]$.\cite{Fert2001} The only temperature-dependent input parameter for the fit process is $\rho^\mathrm{ZnO}(T)$ which was determined independently from the TiN/ZnO/Au reference sample (Fig.\,\ref{fig:fitting}(e)).

The fit curves after Eq.\,(31) of Ref.~\onlinecite{Fert2001} clearly reproduce the measured MR$^\mathrm{max}$, reflecting a fair agreement between model and experiment (red lines in Fig.\,\ref{fig:fitting}(a-d)). From the fits, we extract $l_\mathrm{sf}^\mathrm{ZnO}(T)$ and obtain a value of (10.8$\pm$1.0)\,nm for 2\,K (Fig.\,\ref{fig:fitting}(f)). For increasing temperature, $l_\mathrm{sf}^\mathrm{ZnO}(T)$ stays first nearly constant with a value of (10.7$\pm$0.2)\,nm at 10\,K. For higher temperature, it decreases to (6.2$\pm$0.2)\,nm at 200\,K. We further extract the interface resistance $r_\mathrm{b}(T) \simeq 15\,\mathrm{n}\Omega\mathrm{m}^2$ and the spin selectivity $\gamma(T) \simeq 0.5$, and find both values nearly independent of temperature (see supplemental material). For comparison, we calculate MR$^\mathrm{max}$ for $t_\mathrm{ZnO}=3$\,nm and 10\,nm at 90\,K and obtain 12\% and 3.7\%, respectively, by far exceeding the previously reported values of 1.38\% and 1.12\%.\cite{Ji2009}

The spin diffusion length $l_\mathrm{sf} = \sqrt{\ell \lambda_\mathrm{sf}/6}$ is determined by the transport mean free path $\ell = \sqrt{\langle v^2 \rangle}\tau$ and the spin-flip length $\lambda_\mathrm{sf}=\sqrt{\langle v^2 \rangle}\tau_\mathrm{sf}$.\cite{Fert2001,Bass2007} Here, $\tau_\mathrm{sf}$ is the spin lifetime and $\tau$ the mean free time between collisions. The latter can be derived from the measured resistivity $\rho(T)$ by using the simple Drude relation $\tau = \frac{m^\star}{ne^2\rho}$. For the mean square velocity of the charge carriers we use $\langle v^2 \rangle = \langle v_\mathrm{th}^2 \rangle + \langle v_\mathrm{F}^2 \rangle$, since we are moving from a degenerate Fermi to a classical gas on increasing the temperature. With $\langle v^2_\mathrm{th} \rangle = 3k_\mathrm{B}T/m^\star$ and $\langle v^2_\mathrm{F} \rangle = 2k_\mathrm{B}T_\mathrm{F}/m^\star$, we obtain
\begin{equation}
    \label{eq:tau}
    \tau_\mathrm{sf}^\mathrm{ZnO}(T) = \frac{2 n^\mathrm{ZnO} e^2 \rho^\mathrm{ZnO}(T)}{k_\mathrm{B}(T + \frac{2}{3}T_\mathrm{F})}
                                       (l_{\mathrm{sf}}^\mathrm{ZnO}(T))^2.
\end{equation}
With the experimental values $n^\mathrm{ZnO} \simeq$ 5$\times$10$^{17}\,\mathrm{cm}^{-3}$, $T_\mathrm{F}\!=\!89$\,K, and $\rho^\mathrm{ZnO}(T)$ from Fig.\,\ref{fig:fitting}(e), the spin relaxation rate $1/\tau_\mathrm{sf}^\mathrm{ZnO}$ can be derived. It decreases from (32$\pm$1.7)\,ns$^{-1}$ at 200\,K to (0.49$\pm$0.011)\,ns$^{-1}$ at 10\,K and further to (0.39$\pm$0.0021)\,ns$^{-1}$ at 2\,K (Fig.\,\ref{fig:fitting}(g)). Evidently, this behavior obtained from our all-electrical detection scheme agrees well with the temperature dependence of $1/T_2^\star$ derived from optical TRFR experiments in ZnO thin films\cite{Ghosh2005} in an intermediate temperature range of $10\,\mathrm{K} \leq T \leq 150\,\mathrm{K}$ (Fig.\,\ref{fig:fitting}(g)). For lower temperatures, however, $1/T_2^\star$ increases again while $1/\tau_\mathrm{sf}^\mathrm{ZnO}$ becomes constant below 10\,K.

Two different spin relaxation mechanisms are considered as important in ZnO: the Dzyaloshinsky-Moriya (DM) mechanism due to an anisotropic exchange between localized electronic states and the D'yakonov-Perel' (DP) mechanism due to the reflection asymmetry of the crystal lattice along the $c$-axis.\cite{Harmon2009} For high temperatures ($T>50$\,K), the DP process becomes dominant.\cite{Harmon2009} Its temperature dependence is given by\cite{Harmon2009}
\begin{equation}
    \label{eq:DP}
    \tau_\mathrm{DP}^{-1}(T)=\alpha_\mathrm{DP}\tau(T)T
\end{equation}
with the material-dependent efficiency $\alpha_\mathrm{DP}$. A fit after Eq.\,(\ref{eq:DP}) reproduces our data very well for $T \gtrsim 30$\,K (blue line in Fig.\,\ref{fig:fitting}(g)). We note that the fit yields $\alpha_\mathrm{DP} = 2.0\!\times\!10^6\,\mathrm{K^{-1} ns^{-2}}$ being four orders of magnitude larger than the theory value of $\alpha_\mathrm{DP} = 34.6\,\mathrm{K^{-1} ns^{-2}}$ calculated for bulk ZnO with low carrier concentration.\cite{Harmon2009} We attribute this discrepancy to the low carrier mobility of our ZnO thin films. For $T<50$\,K, however, the DM mechanism is predicted to become dominant, resulting in a nearly $T$-independent spin relaxation rate.\cite{Harmon2009} This behavior was not reported for optical TRFR experiments in ZnO thin films,\cite{Ghosh2005} but is indeed observed in our ZnO-based spin valves for temperatures down to 2\,K (dashed line in Fig.\,\ref{fig:fitting}(g)).

In summary, we create and detect a spin-polarized ensemble of electrons and demonstrate the transport of this spin information across several nanometers in ZnO. We determine the spin diffusion length $l_\mathrm{sf}^\mathrm{ZnO}$ and the spin lifetime $\tau_\mathrm{sf}^\mathrm{ZnO}$ in an all-electrical experiment and obtain large values of $l_\mathrm{sf}^\mathrm{ZnO} = 10.8$\,nm and $\tau_\mathrm{sf}^\mathrm{ZnO} = 2.6$\,ns at 2\,K. The evolution of the measured spin relaxation rates with temperature is consistent with the D'yakonov-Perel' mechanism for $T \gtrsim 30$\,K. For future semiconductor spintronic devices, such all-electrical experiments will be mandatory to extract the relevant spin transport parameters.


\begin{acknowledgments}
We thank A.~Erb for the preparation of the polycrystalline PLD targets, T.~Brenninger for technical support, and M.~S.~Brandt for stimulating discussions. This work was supported by the Deutsche Forschungsgemeinschaft via SPP~1285 (project no.~GR~1132/14) and the German Excellence Initiative via the ``Nanosystems Initiative Munich (NIM)''.
\end{acknowledgments}

\bibliography{znospinvalve}

\providecommand{\noopsort}[1]{}\providecommand{\singleletter}[1]{#1}%
\begin{thebibliography}{30}%
\makeatletter
\providecommand \@ifxundefined [1]{%
 \@ifx{#1\undefined}
}%
\providecommand \@ifnum [1]{%
 \ifnum #1\expandafter \@firstoftwo
 \else \expandafter \@secondoftwo
 \fi
}%
\providecommand \@ifx [1]{%
 \ifx #1\expandafter \@firstoftwo
 \else \expandafter \@secondoftwo
 \fi
}%
\providecommand \natexlab [1]{#1}%
\providecommand \enquote  [1]{``#1''}%
\providecommand \bibnamefont  [1]{#1}%
\providecommand \bibfnamefont [1]{#1}%
\providecommand \citenamefont [1]{#1}%
\providecommand \href@noop [0]{\@secondoftwo}%
\providecommand \href [0]{\begingroup \@sanitize@url \@href}%
\providecommand \@href[1]{\@@startlink{#1}\@@href}%
\providecommand \@@href[1]{\endgroup#1\@@endlink}%
\providecommand \@sanitize@url [0]{\catcode `\\12\catcode `\$12\catcode
  `\&12\catcode `\#12\catcode `\^12\catcode `\_12\catcode `\%12\relax}%
\providecommand \@@startlink[1]{}%
\providecommand \@@endlink[0]{}%
\providecommand \url  [0]{\begingroup\@sanitize@url \@url }%
\providecommand \@url [1]{\endgroup\@href {#1}{\urlprefix }}%
\providecommand \urlprefix  [0]{URL }%
\providecommand \Eprint [0]{\href }%
\providecommand \doibase [0]{http://dx.doi.org/}%
\providecommand \selectlanguage [0]{\@gobble}%
\providecommand \bibinfo  [0]{\@secondoftwo}%
\providecommand \bibfield  [0]{\@secondoftwo}%
\providecommand \translation [1]{[#1]}%
\providecommand \BibitemOpen [0]{}%
\providecommand \bibitemStop [0]{}%
\providecommand \bibitemNoStop [0]{.\EOS\space}%
\providecommand \EOS [0]{\spacefactor3000\relax}%
\providecommand \BibitemShut  [1]{\csname bibitem#1\endcsname}%
\let\auto@bib@innerbib\@empty
\bibitem [{\citenamefont {Prinz}(1998)}]{Prinz1998}%
  \BibitemOpen
  \bibfield  {author} {\bibinfo {author} {\bibfnamefont {G.~A.}\ \bibnamefont
  {Prinz}},\ }\href {\doibase 10.1126/science.282.5394.1660} {\bibfield
  {journal} {\bibinfo  {journal} {Science}\ }\textbf {\bibinfo {volume}
  {282}},\ \bibinfo {pages} {1660} (\bibinfo {year} {1998})}\BibitemShut
  {NoStop}%
\bibitem [{\citenamefont {Fu}\ and\ \citenamefont {Wu}(2008)}]{Fu2008}%
  \BibitemOpen
  \bibfield  {author} {\bibinfo {author} {\bibfnamefont {J.~Y.}\ \bibnamefont
  {Fu}}\ and\ \bibinfo {author} {\bibfnamefont {M.~W.}\ \bibnamefont {Wu}},\
  }\href {\doibase 10.1063/1.3018600} {\bibfield  {journal} {\bibinfo
  {journal} {J. Appl. Phys.}\ }\textbf {\bibinfo {volume} {104}},\ \bibinfo
  {pages} {093712} (\bibinfo {year} {2008})}\BibitemShut {NoStop}%
\bibitem [{\citenamefont {Awschalom}\ and\ \citenamefont
  {Samarth}(2002)}]{Awschalom2002}%
  \BibitemOpen
  \bibfield  {author} {\bibinfo {author} {\bibfnamefont {D.~D.}\ \bibnamefont
  {Awschalom}}\ and\ \bibinfo {author} {\bibfnamefont {N.}~\bibnamefont
  {Samarth}},\ }in\ \href@noop {} {\emph {\bibinfo {booktitle} {Semiconductor
  Spintronics and Quantum Computation}}},\ \bibinfo {editor} {edited by\
  \bibinfo {editor} {\bibfnamefont {D.~D.}\ \bibnamefont {Awschalom}}, \bibinfo
  {editor} {\bibfnamefont {D.}~\bibnamefont {Loss}}, \ and\ \bibinfo {editor}
  {\bibfnamefont {N.}~\bibnamefont {Samarth}}}\ (\bibinfo  {publisher}
  {Springer},\ \bibinfo {address} {Berlin},\ \bibinfo {year} {2002})\ pp.\
  \bibinfo {pages} {147--193}\BibitemShut {NoStop}%
\bibitem [{\citenamefont {Ghosh}\ \emph {et~al.}(2005)\citenamefont {Ghosh},
  \citenamefont {Sih}, \citenamefont {Lau}, \citenamefont {Awschalom},
  \citenamefont {Bae}, \citenamefont {Wang}, \citenamefont {Vaidya},\ and\
  \citenamefont {Chapline}}]{Ghosh2005}%
  \BibitemOpen
  \bibfield  {author} {\bibinfo {author} {\bibfnamefont {S.}~\bibnamefont
  {Ghosh}}, \bibinfo {author} {\bibfnamefont {V.}~\bibnamefont {Sih}}, \bibinfo
  {author} {\bibfnamefont {W.~H.}\ \bibnamefont {Lau}}, \bibinfo {author}
  {\bibfnamefont {D.~D.}\ \bibnamefont {Awschalom}}, \bibinfo {author}
  {\bibfnamefont {S.}~\bibnamefont {Bae}}, \bibinfo {author} {\bibfnamefont
  {S.}~\bibnamefont {Wang}}, \bibinfo {author} {\bibfnamefont {S.}~\bibnamefont
  {Vaidya}}, \ and\ \bibinfo {author} {\bibfnamefont {G.}~\bibnamefont
  {Chapline}},\ }\href {\doibase 10.1063/1.1946204} {\bibfield  {journal}
  {\bibinfo  {journal} {Appl. Phys. Lett.}\ }\textbf {\bibinfo {volume} {86}},\
  \bibinfo {pages} {232507} (\bibinfo {year} {2005})}\BibitemShut {NoStop}%
\bibitem [{\citenamefont {Liu}\ \emph {et~al.}(2007)\citenamefont {Liu},
  \citenamefont {Whitaker}, \citenamefont {Smith}, \citenamefont {Kittilstved},
  \citenamefont {Robinson},\ and\ \citenamefont {Gamelin}}]{Liu2007}%
  \BibitemOpen
  \bibfield  {author} {\bibinfo {author} {\bibfnamefont {W.~K.}\ \bibnamefont
  {Liu}}, \bibinfo {author} {\bibfnamefont {K.~M.}\ \bibnamefont {Whitaker}},
  \bibinfo {author} {\bibfnamefont {A.~L.}\ \bibnamefont {Smith}}, \bibinfo
  {author} {\bibfnamefont {K.~R.}\ \bibnamefont {Kittilstved}}, \bibinfo
  {author} {\bibfnamefont {B.~H.}\ \bibnamefont {Robinson}}, \ and\ \bibinfo
  {author} {\bibfnamefont {D.~R.}\ \bibnamefont {Gamelin}},\ }\href {\doibase
  10.1103/PhysRevLett.98.186804} {\bibfield  {journal} {\bibinfo  {journal}
  {Phys. Rev. Lett.}\ }\textbf {\bibinfo {volume} {98}},\ \bibinfo {pages}
  {186804} (\bibinfo {year} {2007})}\BibitemShut {NoStop}%
\bibitem [{\citenamefont {Jan{\ss}en}\ \emph {et~al.}(2008)\citenamefont
  {Jan{\ss}en}, \citenamefont {Whitaker}, \citenamefont {Gamelin},\ and\
  \citenamefont {Bratschitsch}}]{Bratschitsch2008}%
  \BibitemOpen
  \bibfield  {author} {\bibinfo {author} {\bibfnamefont {N.}~\bibnamefont
  {Jan{\ss}en}}, \bibinfo {author} {\bibfnamefont {K.~M.}\ \bibnamefont
  {Whitaker}}, \bibinfo {author} {\bibfnamefont {D.~R.}\ \bibnamefont
  {Gamelin}}, \ and\ \bibinfo {author} {\bibfnamefont {R.}~\bibnamefont
  {Bratschitsch}},\ }\href {\doibase 10.1021/nl801057q} {\bibfield  {journal}
  {\bibinfo  {journal} {Nano Lett.}\ }\textbf {\bibinfo {volume} {8}},\
  \bibinfo {pages} {1991} (\bibinfo {year} {2008})}\BibitemShut {NoStop}%
\bibitem [{\citenamefont {Weier}(2010)}]{Weier2010}%
  \BibitemOpen
  \bibfield  {author} {\bibinfo {author} {\bibfnamefont {C.}~\bibnamefont
  {Weier}},\ }\emph {\bibinfo {title} {Optische Untersuchung der Spindynamik
  und der elektrischen Spininjektion in Zinkoxid}},\ \href@noop {} {\bibinfo
  {type} {Diploma thesis}},\ \bibinfo  {school} {RWTH Aachen}, \bibinfo
  {address} {II. Physikalisches Institut, Lehrstuhl A} (\bibinfo {year}
  {2010})\BibitemShut {NoStop}%
\bibitem [{\citenamefont {Opel}(2012)}]{Opel2012}%
  \BibitemOpen
  \bibfield  {author} {\bibinfo {author} {\bibfnamefont {M.}~\bibnamefont
  {Opel}},\ }\href {\doibase 10.1088/0022-3727/45/3/033001} {\bibfield
  {journal} {\bibinfo  {journal} {J.\ Phys.\ D:\ Appl.\ Phys.}\ }\textbf
  {\bibinfo {volume} {45}},\ \bibinfo {pages} {033001} (\bibinfo {year}
  {2012})}\BibitemShut {NoStop}%
\bibitem [{\citenamefont {Chen}\ \emph {et~al.}(2002)\citenamefont {Chen},
  \citenamefont {Ren}, \citenamefont {Ji}, \citenamefont {Fang}, \citenamefont
  {Chen}, \citenamefont {Xiao}, \citenamefont {Xie}, \citenamefont {Liu},\ and\
  \citenamefont {Mei}}]{Chen2002}%
  \BibitemOpen
  \bibfield  {author} {\bibinfo {author} {\bibfnamefont {Y.}~\bibnamefont
  {Chen}}, \bibinfo {author} {\bibfnamefont {M.}~\bibnamefont {Ren}}, \bibinfo
  {author} {\bibfnamefont {G.}~\bibnamefont {Ji}}, \bibinfo {author}
  {\bibfnamefont {J.}~\bibnamefont {Fang}}, \bibinfo {author} {\bibfnamefont
  {J.}~\bibnamefont {Chen}}, \bibinfo {author} {\bibfnamefont {S.}~\bibnamefont
  {Xiao}}, \bibinfo {author} {\bibfnamefont {S.}~\bibnamefont {Xie}}, \bibinfo
  {author} {\bibfnamefont {Y.}~\bibnamefont {Liu}}, \ and\ \bibinfo {author}
  {\bibfnamefont {L.}~\bibnamefont {Mei}},\ }\href {\doibase
  10.1016/S0375-9601(02)01204-5} {\bibfield  {journal} {\bibinfo  {journal}
  {Physics Letters A}\ }\textbf {\bibinfo {volume} {303}},\ \bibinfo {pages}
  {91 } (\bibinfo {year} {2002})}\BibitemShut {NoStop}%
\bibitem [{\citenamefont {Ji}\ \emph {et~al.}(2009)\citenamefont {Ji},
  \citenamefont {Zhang}, \citenamefont {Chen}, \citenamefont {Yan},
  \citenamefont {Liu},\ and\ \citenamefont {Mei}}]{Ji2009}%
  \BibitemOpen
  \bibfield  {author} {\bibinfo {author} {\bibfnamefont {G.}~\bibnamefont
  {Ji}}, \bibinfo {author} {\bibfnamefont {Z.}~\bibnamefont {Zhang}}, \bibinfo
  {author} {\bibfnamefont {Y.}~\bibnamefont {Chen}}, \bibinfo {author}
  {\bibfnamefont {S.}~\bibnamefont {Yan}}, \bibinfo {author} {\bibfnamefont
  {Y.}~\bibnamefont {Liu}}, \ and\ \bibinfo {author} {\bibfnamefont
  {L.}~\bibnamefont {Mei}},\ }\href {\doibase 10.1016/S1006-7191(08)60083-6}
  {\bibfield  {journal} {\bibinfo  {journal} {Acta Metallurgica Sinica (English
  Letters)}\ }\textbf {\bibinfo {volume} {22}},\ \bibinfo {pages} {153}
  (\bibinfo {year} {2009})}\BibitemShut {NoStop}%
\bibitem [{\citenamefont {Shimazawa}\ \emph {et~al.}(2010)\citenamefont
  {Shimazawa}, \citenamefont {Tsuchiya}, \citenamefont {Mizuno}, \citenamefont
  {Hara}, \citenamefont {Chou}, \citenamefont {Miyauchi}, \citenamefont
  {Machita}, \citenamefont {Ayukawa}, \citenamefont {Ichiki},\ and\
  \citenamefont {Noguchi}}]{Shimazawa2010}%
  \BibitemOpen
  \bibfield  {author} {\bibinfo {author} {\bibfnamefont {K.}~\bibnamefont
  {Shimazawa}}, \bibinfo {author} {\bibfnamefont {Y.}~\bibnamefont {Tsuchiya}},
  \bibinfo {author} {\bibfnamefont {T.}~\bibnamefont {Mizuno}}, \bibinfo
  {author} {\bibfnamefont {S.}~\bibnamefont {Hara}}, \bibinfo {author}
  {\bibfnamefont {T.}~\bibnamefont {Chou}}, \bibinfo {author} {\bibfnamefont
  {D.}~\bibnamefont {Miyauchi}}, \bibinfo {author} {\bibfnamefont
  {T.}~\bibnamefont {Machita}}, \bibinfo {author} {\bibfnamefont
  {T.}~\bibnamefont {Ayukawa}}, \bibinfo {author} {\bibfnamefont
  {T.}~\bibnamefont {Ichiki}}, \ and\ \bibinfo {author} {\bibfnamefont
  {K.}~\bibnamefont {Noguchi}},\ }\href {\doibase 10.1109/TMAG.2010.2042574}
  {\bibfield  {journal} {\bibinfo  {journal} {IEEE Trans. Mag.}\ }\textbf
  {\bibinfo {volume} {46}},\ \bibinfo {pages} {1487} (\bibinfo {year}
  {2010})}\BibitemShut {NoStop}%
\bibitem [{\citenamefont {Schmidt}\ \emph {et~al.}(2000)\citenamefont
  {Schmidt}, \citenamefont {Ferrand}, \citenamefont {Molenkamp}, \citenamefont
  {Filip},\ and\ \citenamefont {van Wees}}]{Schmidt2000}%
  \BibitemOpen
  \bibfield  {author} {\bibinfo {author} {\bibfnamefont {G.}~\bibnamefont
  {Schmidt}}, \bibinfo {author} {\bibfnamefont {D.}~\bibnamefont {Ferrand}},
  \bibinfo {author} {\bibfnamefont {L.~W.}\ \bibnamefont {Molenkamp}}, \bibinfo
  {author} {\bibfnamefont {A.~T.}\ \bibnamefont {Filip}}, \ and\ \bibinfo
  {author} {\bibfnamefont {B.~J.}\ \bibnamefont {van Wees}},\ }\href {\doibase
  10.1103/PhysRevB.62.R4790} {\bibfield  {journal} {\bibinfo  {journal} {Phys.
  Rev. B}\ }\textbf {\bibinfo {volume} {62}},\ \bibinfo {pages} {R4790}
  (\bibinfo {year} {2000})}\BibitemShut {NoStop}%
\bibitem [{\citenamefont {Baibich}\ \emph {et~al.}(1988)\citenamefont
  {Baibich}, \citenamefont {Broto}, \citenamefont {Fert}, \citenamefont {van
  Dau}, \citenamefont {Petroff}, \citenamefont {Etienne}, \citenamefont
  {Creuzet}, \citenamefont {Friederich},\ and\ \citenamefont
  {Chazelas}}]{Baibich1988}%
  \BibitemOpen
  \bibfield  {author} {\bibinfo {author} {\bibfnamefont {M.~N.}\ \bibnamefont
  {Baibich}}, \bibinfo {author} {\bibfnamefont {J.~M.}\ \bibnamefont {Broto}},
  \bibinfo {author} {\bibfnamefont {A.}~\bibnamefont {Fert}}, \bibinfo {author}
  {\bibfnamefont {F.~N.}\ \bibnamefont {van Dau}}, \bibinfo {author}
  {\bibfnamefont {F.}~\bibnamefont {Petroff}}, \bibinfo {author} {\bibfnamefont
  {P.}~\bibnamefont {Etienne}}, \bibinfo {author} {\bibfnamefont
  {G.}~\bibnamefont {Creuzet}}, \bibinfo {author} {\bibfnamefont
  {A.}~\bibnamefont {Friederich}}, \ and\ \bibinfo {author} {\bibfnamefont
  {J.}~\bibnamefont {Chazelas}},\ }\href {\doibase 10.1103/PhysRevLett.61.2472}
  {\bibfield  {journal} {\bibinfo  {journal} {Phys. Rev. Lett.}\ }\textbf
  {\bibinfo {volume} {61}},\ \bibinfo {pages} {2472} (\bibinfo {year}
  {1988})}\BibitemShut {NoStop}%
\bibitem [{\citenamefont {Binasch}\ \emph {et~al.}(1989)\citenamefont
  {Binasch}, \citenamefont {Gr{\"u}nberg}, \citenamefont {Saurenbach},\ and\
  \citenamefont {Zinn}}]{Binasch1989}%
  \BibitemOpen
  \bibfield  {author} {\bibinfo {author} {\bibfnamefont {G.}~\bibnamefont
  {Binasch}}, \bibinfo {author} {\bibfnamefont {P.}~\bibnamefont
  {Gr{\"u}nberg}}, \bibinfo {author} {\bibfnamefont {F.}~\bibnamefont
  {Saurenbach}}, \ and\ \bibinfo {author} {\bibfnamefont {W.}~\bibnamefont
  {Zinn}},\ }\href {\doibase 10.1103/PhysRevB.39.4828} {\bibfield  {journal}
  {\bibinfo  {journal} {Phys. Rev. B}\ }\textbf {\bibinfo {volume} {39}},\
  \bibinfo {pages} {4828} (\bibinfo {year} {1989})}\BibitemShut {NoStop}%
\bibitem [{\citenamefont {Gijs}\ and\ \citenamefont {Bauer}(1997)}]{Gijs1997}%
  \BibitemOpen
  \bibfield  {author} {\bibinfo {author} {\bibfnamefont {M.~A.~M.}\
  \bibnamefont {Gijs}}\ and\ \bibinfo {author} {\bibfnamefont {G.~E.~W.}\
  \bibnamefont {Bauer}},\ }\href {\doibase 10.1080/00018739700101518}
  {\bibfield  {journal} {\bibinfo  {journal} {Adv. Phys.}\ }\textbf {\bibinfo
  {volume} {46}},\ \bibinfo {pages} {285} (\bibinfo {year} {1997})}\BibitemShut
  {NoStop}%
\bibitem [{\citenamefont {Gross}\ \emph {et~al.}(2000)\citenamefont {Gross},
  \citenamefont {Klein}, \citenamefont {Wiedenhorst}, \citenamefont
  {H{\"o}fener}, \citenamefont {Schoop}, \citenamefont {Philipp}, \citenamefont
  {Schonecke}, \citenamefont {Herbstritt}, \citenamefont {Alff}, \citenamefont
  {Lu}, \citenamefont {Marx}, \citenamefont {S.Schymon}, \citenamefont
  {Thienhaus},\ and\ \citenamefont {Mader}}]{Gross2000}%
  \BibitemOpen
  \bibfield  {author} {\bibinfo {author} {\bibfnamefont {R.}~\bibnamefont
  {Gross}}, \bibinfo {author} {\bibfnamefont {J.}~\bibnamefont {Klein}},
  \bibinfo {author} {\bibfnamefont {B.}~\bibnamefont {Wiedenhorst}}, \bibinfo
  {author} {\bibfnamefont {C.}~\bibnamefont {H{\"o}fener}}, \bibinfo {author}
  {\bibfnamefont {U.}~\bibnamefont {Schoop}}, \bibinfo {author} {\bibfnamefont
  {J.~B.}\ \bibnamefont {Philipp}}, \bibinfo {author} {\bibfnamefont
  {M.}~\bibnamefont {Schonecke}}, \bibinfo {author} {\bibfnamefont
  {F.}~\bibnamefont {Herbstritt}}, \bibinfo {author} {\bibfnamefont
  {L.}~\bibnamefont {Alff}}, \bibinfo {author} {\bibfnamefont {Y.}~\bibnamefont
  {Lu}}, \bibinfo {author} {\bibfnamefont {A.}~\bibnamefont {Marx}}, \bibinfo
  {author} {\bibnamefont {S.Schymon}}, \bibinfo {author} {\bibfnamefont
  {S.}~\bibnamefont {Thienhaus}}, \ and\ \bibinfo {author} {\bibfnamefont
  {W.}~\bibnamefont {Mader}},\ }\href {\doibase 10.1117/12.397845} {\bibfield
  {journal} {\bibinfo  {journal} {Proc. SPIE}\ }\textbf {\bibinfo {volume}
  {4058}},\ \bibinfo {pages} {278} (\bibinfo {year} {2000})}\BibitemShut
  {NoStop}%
\bibitem [{\citenamefont {Martienssen}\ and\ \citenamefont
  {Warlimont}(2005)}]{Martienssen2005}%
  \BibitemOpen
  \bibfield  {author} {\bibinfo {author} {\bibfnamefont {W.}~\bibnamefont
  {Martienssen}}\ and\ \bibinfo {author} {\bibfnamefont {H.}~\bibnamefont
  {Warlimont}},\ }\href@noop {} {\emph {\bibinfo {title} {Springer Handbook of
  Condensed Matter and Materials Data}}}\ (\bibinfo  {publisher} {Springer},\
  \bibinfo {address} {Heidelberg},\ \bibinfo {year} {2005})\BibitemShut
  {NoStop}%
\bibitem [{\citenamefont {Talyansky}\ \emph {et~al.}(1999)\citenamefont
  {Talyansky}, \citenamefont {Choopun}, \citenamefont {Downes}, \citenamefont
  {Sharma}, \citenamefont {Venkatesan}, \citenamefont {Li}, \citenamefont
  {Salamanca-Riba}, \citenamefont {Wood}, \citenamefont {Lareau},\ and\
  \citenamefont {Jones}}]{Talyansky1999}%
  \BibitemOpen
  \bibfield  {author} {\bibinfo {author} {\bibfnamefont {V.}~\bibnamefont
  {Talyansky}}, \bibinfo {author} {\bibfnamefont {S.}~\bibnamefont {Choopun}},
  \bibinfo {author} {\bibfnamefont {M.~J.}\ \bibnamefont {Downes}}, \bibinfo
  {author} {\bibfnamefont {R.~P.}\ \bibnamefont {Sharma}}, \bibinfo {author}
  {\bibfnamefont {T.}~\bibnamefont {Venkatesan}}, \bibinfo {author}
  {\bibfnamefont {Y.~X.}\ \bibnamefont {Li}}, \bibinfo {author} {\bibfnamefont
  {L.~G.}\ \bibnamefont {Salamanca-Riba}}, \bibinfo {author} {\bibfnamefont
  {M.~C.}\ \bibnamefont {Wood}}, \bibinfo {author} {\bibfnamefont {R.~T.}\
  \bibnamefont {Lareau}}, \ and\ \bibinfo {author} {\bibfnamefont {K.~A.}\
  \bibnamefont {Jones}},\ }\href {\doibase 10.1557/JMR.1999.0446} {\bibfield
  {journal} {\bibinfo  {journal} {J. Mater. Res.}\ }\textbf {\bibinfo {volume}
  {14}},\ \bibinfo {pages} {3298} (\bibinfo {year} {1999})}\BibitemShut
  {NoStop}%
\bibitem [{\citenamefont {Ago}\ \emph {et~al.}(2010)\citenamefont {Ago},
  \citenamefont {Ito}, \citenamefont {Mizuta}, \citenamefont {Yoshida},
  \citenamefont {Hu}, \citenamefont {Orofeo}, \citenamefont {Tsuji},
  \citenamefont {i.~Ikeda},\ and\ \citenamefont {Mizuno}}]{Ago2010}%
  \BibitemOpen
  \bibfield  {author} {\bibinfo {author} {\bibfnamefont {H.}~\bibnamefont
  {Ago}}, \bibinfo {author} {\bibfnamefont {Y.}~\bibnamefont {Ito}}, \bibinfo
  {author} {\bibfnamefont {N.}~\bibnamefont {Mizuta}}, \bibinfo {author}
  {\bibfnamefont {K.}~\bibnamefont {Yoshida}}, \bibinfo {author} {\bibfnamefont
  {B.}~\bibnamefont {Hu}}, \bibinfo {author} {\bibfnamefont {C.~M.}\
  \bibnamefont {Orofeo}}, \bibinfo {author} {\bibfnamefont {M.}~\bibnamefont
  {Tsuji}}, \bibinfo {author} {\bibfnamefont {K.}~\bibnamefont {i.~Ikeda}}, \
  and\ \bibinfo {author} {\bibfnamefont {S.}~\bibnamefont {Mizuno}},\ }\href
  {\doibase 10.1021/nn102519b} {\bibfield  {journal} {\bibinfo  {journal} {ACS
  Nano}\ }\textbf {\bibinfo {volume} {4}},\ \bibinfo {pages} {7407} (\bibinfo
  {year} {2010})}\BibitemShut {NoStop}%
\bibitem [{\citenamefont {Chen}\ \emph {et~al.}(1998)\citenamefont {Chen},
  \citenamefont {Bagnall}, \citenamefont {Koh}, \citenamefont {Park},
  \citenamefont {Hiraga}, \citenamefont {Zhu},\ and\ \citenamefont
  {Yao}}]{Chen1998}%
  \BibitemOpen
  \bibfield  {author} {\bibinfo {author} {\bibfnamefont {Y.}~\bibnamefont
  {Chen}}, \bibinfo {author} {\bibfnamefont {D.~M.}\ \bibnamefont {Bagnall}},
  \bibinfo {author} {\bibfnamefont {H.-J.}\ \bibnamefont {Koh}}, \bibinfo
  {author} {\bibfnamefont {K.-T.}\ \bibnamefont {Park}}, \bibinfo {author}
  {\bibfnamefont {K.}~\bibnamefont {Hiraga}}, \bibinfo {author} {\bibfnamefont
  {Z.}~\bibnamefont {Zhu}}, \ and\ \bibinfo {author} {\bibfnamefont
  {T.}~\bibnamefont {Yao}},\ }\href {\doibase 10.1063/1.368595} {\bibfield
  {journal} {\bibinfo  {journal} {J. Appl. Phys.}\ }\textbf {\bibinfo {volume}
  {84}},\ \bibinfo {pages} {3912} (\bibinfo {year} {1998})}\BibitemShut
  {NoStop}%
\bibitem [{\citenamefont {K{\"a}stle}\ \emph {et~al.}(2002)\citenamefont
  {K{\"a}stle}, \citenamefont {Boyen}, \citenamefont {Koslowski}, \citenamefont
  {Plettl}, \citenamefont {Weigl},\ and\ \citenamefont {Ziemann}}]{Kastle2002}%
  \BibitemOpen
  \bibfield  {author} {\bibinfo {author} {\bibfnamefont {G.}~\bibnamefont
  {K{\"a}stle}}, \bibinfo {author} {\bibfnamefont {H.}~\bibnamefont {Boyen}},
  \bibinfo {author} {\bibfnamefont {B.}~\bibnamefont {Koslowski}}, \bibinfo
  {author} {\bibfnamefont {A.}~\bibnamefont {Plettl}}, \bibinfo {author}
  {\bibfnamefont {F.}~\bibnamefont {Weigl}}, \ and\ \bibinfo {author}
  {\bibfnamefont {P.}~\bibnamefont {Ziemann}},\ }\href {\doibase
  10.1016/S0039-6028(01)01685-5} {\bibfield  {journal} {\bibinfo  {journal}
  {Surf. Sci.}\ }\textbf {\bibinfo {volume} {498}},\ \bibinfo {pages} {168}
  (\bibinfo {year} {2002})}\BibitemShut {NoStop}%
\bibitem [{Note1()}]{Note1}%
  \BibitemOpen
  \bibinfo {note} {The linear behavior is expected in the tunneling regime. For
  higher currents, heating leads to non-linear
  $I$-$V$-characteristics.}\BibitemShut {Stop}%
\bibitem [{Note2()}]{Note2}%
  \BibitemOpen
  \bibinfo {note} {This statement is further confirmed by angle-resolved
  measurements, see supplemental material.}\BibitemShut {Stop}%
\bibitem [{\citenamefont {Valet}\ and\ \citenamefont {Fert}(1993)}]{Valet1993}%
  \BibitemOpen
  \bibfield  {author} {\bibinfo {author} {\bibfnamefont {T.}~\bibnamefont
  {Valet}}\ and\ \bibinfo {author} {\bibfnamefont {A.}~\bibnamefont {Fert}},\
  }\href {\doibase 10.1103/PhysRevB.48.7099} {\bibfield  {journal} {\bibinfo
  {journal} {Phys. Rev. B}\ }\textbf {\bibinfo {volume} {48}},\ \bibinfo
  {pages} {7099} (\bibinfo {year} {1993})}\BibitemShut {NoStop}%
\bibitem [{\citenamefont {Fert}\ and\ \citenamefont
  {Jaffr\`es}(2001)}]{Fert2001}%
  \BibitemOpen
  \bibfield  {author} {\bibinfo {author} {\bibfnamefont {A.}~\bibnamefont
  {Fert}}\ and\ \bibinfo {author} {\bibfnamefont {H.}~\bibnamefont
  {Jaffr\`es}},\ }\href {\doibase 10.1103/PhysRevB.64.184420} {\bibfield
  {journal} {\bibinfo  {journal} {Phys. Rev. B}\ }\textbf {\bibinfo {volume}
  {64}},\ \bibinfo {pages} {184420} (\bibinfo {year} {2001})}\BibitemShut
  {NoStop}%
\bibitem [{\citenamefont {Moreau}\ \emph {et~al.}(2007)\citenamefont {Moreau},
  \citenamefont {Moraru}, \citenamefont {Birge},\ and\ \citenamefont
  {Pratt}}]{Moreau2007}%
  \BibitemOpen
  \bibfield  {author} {\bibinfo {author} {\bibfnamefont {C.~E.}\ \bibnamefont
  {Moreau}}, \bibinfo {author} {\bibfnamefont {I.~C.}\ \bibnamefont {Moraru}},
  \bibinfo {author} {\bibfnamefont {N.~O.}\ \bibnamefont {Birge}}, \ and\
  \bibinfo {author} {\bibfnamefont {W.~P.}\ \bibnamefont {Pratt}},\ }\href
  {\doibase 10.1063/1.2424437} {\bibfield  {journal} {\bibinfo  {journal}
  {Applied Physics Letters}\ }\textbf {\bibinfo {volume} {90}},\ \bibinfo {eid}
  {012101} (\bibinfo {year} {2007})}\BibitemShut {NoStop}%
\bibitem [{\citenamefont {Bass}\ and\ \citenamefont {Pratt}(1999)}]{Bass1999}%
  \BibitemOpen
  \bibfield  {author} {\bibinfo {author} {\bibfnamefont {J.}~\bibnamefont
  {Bass}}\ and\ \bibinfo {author} {\bibfnamefont {W.~P.}\ \bibnamefont
  {Pratt}},\ }\href {\doibase 10.1016/S0304-8853(99)00316-9} {\bibfield
  {journal} {\bibinfo  {journal} {J. Magn. Magn. Mater.}\ }\textbf {\bibinfo
  {volume} {200}},\ \bibinfo {pages} {274} (\bibinfo {year}
  {1999})}\BibitemShut {NoStop}%
\bibitem [{\citenamefont {Piraux}\ \emph {et~al.}(1998)\citenamefont {Piraux},
  \citenamefont {Dubois}, \citenamefont {Fert},\ and\ \citenamefont
  {Belliard}}]{Piraux1998}%
  \BibitemOpen
  \bibfield  {author} {\bibinfo {author} {\bibfnamefont {L.}~\bibnamefont
  {Piraux}}, \bibinfo {author} {\bibfnamefont {S.}~\bibnamefont {Dubois}},
  \bibinfo {author} {\bibfnamefont {A.}~\bibnamefont {Fert}}, \ and\ \bibinfo
  {author} {\bibfnamefont {L.}~\bibnamefont {Belliard}},\ }\href {\doibase
  10.1007/s100510050398} {\bibfield  {journal} {\bibinfo  {journal} {Eur. Phys.
  J. B}\ }\textbf {\bibinfo {volume} {4}},\ \bibinfo {pages} {413} (\bibinfo
  {year} {1998})}\BibitemShut {NoStop}%
\bibitem [{\citenamefont {Bass}\ and\ \citenamefont {Pratt}(2007)}]{Bass2007}%
  \BibitemOpen
  \bibfield  {author} {\bibinfo {author} {\bibfnamefont {J.}~\bibnamefont
  {Bass}}\ and\ \bibinfo {author} {\bibfnamefont {W.~P.}\ \bibnamefont
  {Pratt}},\ }\href {\doibase 10.1088/0953-8984/19/18/183201} {\bibfield
  {journal} {\bibinfo  {journal} {J. Phys.: Condens. Matter}\ }\textbf
  {\bibinfo {volume} {19}},\ \bibinfo {pages} {183201} (\bibinfo {year}
  {2007})}\BibitemShut {NoStop}%
\bibitem [{\citenamefont {Harmon}, \citenamefont {Putikka},\ and\ \citenamefont
  {Joynt}(2009)}]{Harmon2009}%
  \BibitemOpen
  \bibfield  {author} {\bibinfo {author} {\bibfnamefont {N.~J.}\ \bibnamefont
  {Harmon}}, \bibinfo {author} {\bibfnamefont {W.~O.}\ \bibnamefont {Putikka}},
  \ and\ \bibinfo {author} {\bibfnamefont {R.}~\bibnamefont {Joynt}},\ }\href
  {\doibase 10.1103/PhysRevB.79.115204} {\bibfield  {journal} {\bibinfo
  {journal} {Phys. Rev. B}\ }\textbf {\bibinfo {volume} {79}},\ \bibinfo
  {pages} {115204} (\bibinfo {year} {2009})}\BibitemShut {NoStop}%
\end{thebibliography}%

\clearpage

\section*{Supplement I: \newline
Angle-resolved magnetoresistance}

\begin{figure}[b]
    \includegraphics[width=\columnwidth]{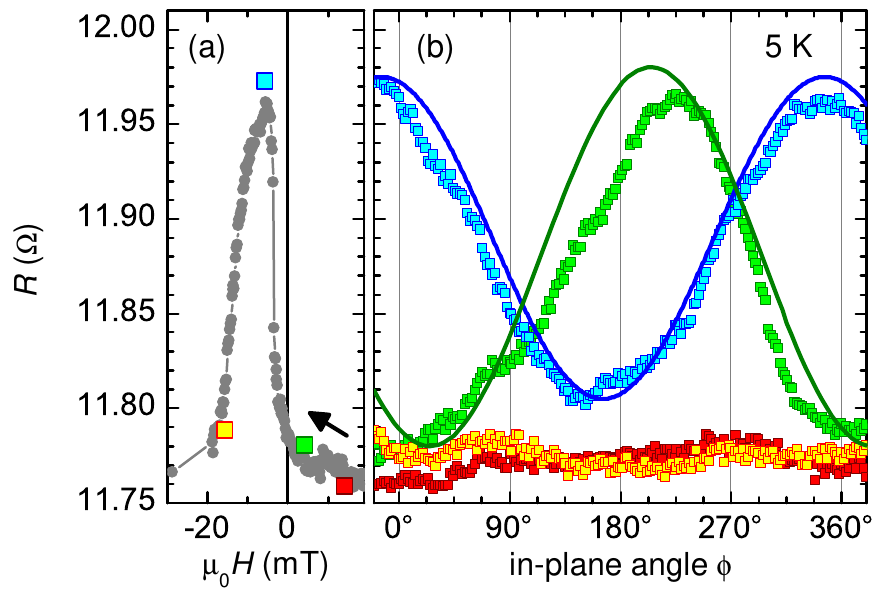}
    \caption{\label{fig:AMR} (a) Magnetoresistance for a spin valve device with an area of $400\,\mu\mathrm{m}^2$. The data were taken for a ZnO thickness of 20\,nm at 5\,K as a function of the in-plane magnetic field $H$ during down sweep (grey). At different fields of 14.3\,mT (red square), 4.3\,mT (green square), $-5.7$\,mT (blue square), and $-15.7$\,mT (yellow square), the field sweep was paused. (b) At these fields, the sample was rotated by $360^\circ$ at constant in-plane magnetic field (squares of the respective color). For details see text. The green and blue lines are fits with a shifted cosine function. All data were taken from the same sample as in Fig.~2 of the main text.}
\end{figure}

To provide further evidence for the spin valve behavior of our samples, we performed angle-resolved measurements of the magnetoresistance at 5\,K (Fig.~\ref{fig:AMR}), according to the following procedure: (i) We first applied a high magnetic field $\mu_0 H$ at a fixed in-plane angle of $\phi \equiv 0^\circ$ such that the magnetizations of both Ni and Co electrodes are aligned parallel. (ii) Second, we swept down $\mu_0 H$ at $\phi = 0^\circ$ while recording the resistance $R$ of the device. (iii) Third, we paused the field sweep at $14.3$\,mT (red square in Fig.~\ref{fig:AMR}(a)), i.e.~above the coercive fields $H_\mathrm{c}^\mathrm{Ni},H_\mathrm{c}^\mathrm{Co}$ of both Ni and Co, and rotated the sample with respect to the field direction such that $R(\phi)$ could be recorded for $0^\circ \leq \phi \leq 360^\circ$ at constant in-plane magnetic field (red squares in Fig.~\ref{fig:AMR}(b)). We then resumed the field sweep of step (ii) and successively repeated step (iii) at $4.3$\,mT between $H_\mathrm{c}^\mathrm{Co}$ and $H_\mathrm{c}^\mathrm{Ni}$ (green squares), at $-5.7$\,mT between $-H_\mathrm{c}^\mathrm{Ni}$ and $-H_\mathrm{c}^\mathrm{Co}$ (blue squares), and at $-15.7$\,mT below $-H_\mathrm{c}^\mathrm{Ni}$ and $-H_\mathrm{c}^\mathrm{Co}$ (yellow squares).

Above or below the coercivities of both Ni and Co electrodes, the resistance $R$ is low and does not vary significantly with $\phi$ (red and yellow symbols in Fig.~\ref{fig:AMR}(b)). This demonstrates that the magnetization directions in both ferromagnetic electrodes always stay aligned parallel and therefore follow the applied external field direction. For fields in between the coercivities of Ni and Co, however, the situation is different. $R$ displays a variation between its previously observed low value and a high resistive state with the same value as in the field sweep (Fig.~\ref{fig:AMR}(a)), following a $\cos\phi$ behavior (green and blue symbols in Fig.~\ref{fig:AMR}(b)). This observation is in agreement with the assumption that the magnetization of the Co electrode (with lower coercivity) follows the external field direction whereas the magnetization of Ni (with higher coercivity) stays unaffected. This behavior is expected for a spin valve device as is the resulting $R\!\propto\!\cos\phi$ dependence (green and blue lines in Fig.~\ref{fig:AMR}(b)). It is in clear contradiction to an anisotropic magnetoresistance (AMR) effect of the ferromagnetic electrodes which would result in a $R\!\propto\!\cos^2\phi$ dependence.

\clearpage

\section*{Supplement II: \newline
Interface resistance and spin selectivity}

\begin{figure}[h]
    \includegraphics[width=\columnwidth]{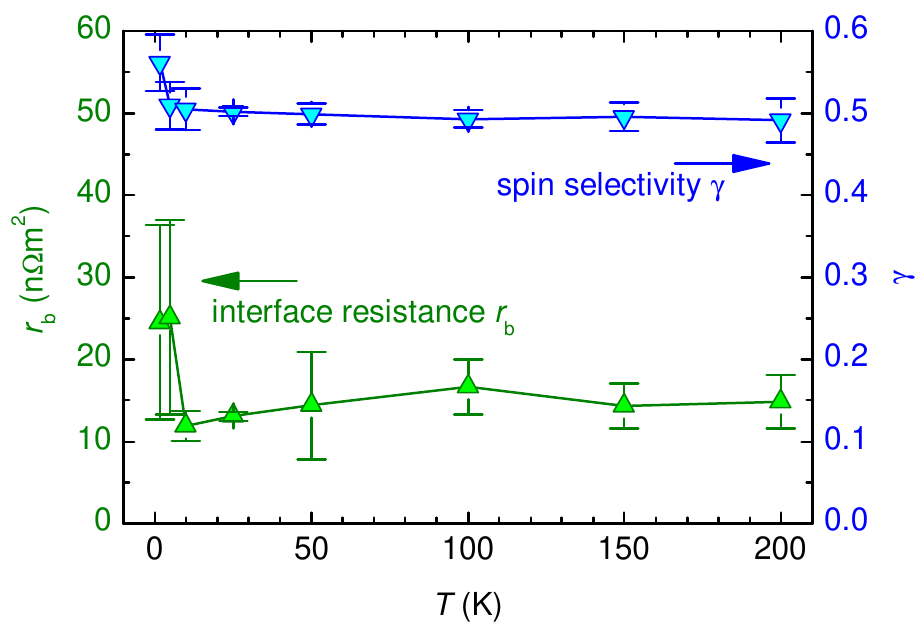}
    \caption{\label{fig:rb} Interface resistance area product $r_\mathrm{b}(T)$ (up green triangles, left scale) and spin selectivity $\gamma(T)$ (down blue triangles, right scale) as a function of the temperature $T$. The data were obtained from the fits in Fig.~3(a--d) following the Valet-Fert approach described in Refs.~[24,25].}
\end{figure}

\end{document}